\documentclass[11 pt,superscriptaddress,showkeys,preprint]{revtex4}
\usepackage{mathrsfs}
\usepackage{amsfonts}
\usepackage[mathscr]{eucal}
\usepackage{amssymb}
\usepackage{amsmath}
\usepackage{dcolumn}
\usepackage{bm}
\usepackage{xcolor}
\usepackage{graphicx}
\begin{document}

\newcommand\ppcf{Plasma Phys. Control. Fusion }
\newcommand\pop{Phys. Plasmas }


\title{Landau Damping of Geodesic Acoustic Mode in Toroidally Rotating Tokamaks}
\author{Haijun Ren}
\email{hjren@ustc.edu.cn}
\affiliation{The Collaborative Innovation Center for Advanced Fusion Energy and Plasma Science, and Department of Modern Physics, University of Science and Technology of China, Hefei 230026, P. R. China}

\author{Jintao Cao}
\affiliation{Bejing National Laboratory for Condensed Matter Physics and CAS Key Laboratory of Soft Matter Physics, Institute of Physics, Chinese Academy of Sciences, Beijing 100190, P. R. China}

\begin{abstract}
Geodesic acoustic mode (GAM) is analyzed by using modified gyro-kinetic (MGK) equation applicable to low-frequency microinstabilities in a rotating axisymmetric plasma. Dispersion relation of GAM in the presence of arbitrary Mach number is analytically derived. Toroidal rotation plays the same effects on the GAM regardless of the orientation of equilibrium flow. It is shown that the toroidal Mach number $M$ increases the GAM frequency and dramatically decreases the Landau damping rate. The valid of classical gyro-kinetic (CGK) equation is also examined. For zero electron temperature, CGK is identical with MGK. For non-zero electron temperature, CGK gives the same real frequency of GAM as MGK but induces an instability with a growth rate proportional to $M^3/q$, where $q$ is the safety factor.
\end{abstract}


\keywords{Geodesic Acoustic Mode, Toroidal Rotation, Collisionless}
\date{\today}

\maketitle

\section{Introduction}
Geodesic acoustic mode (GAM)\cite{winsor}, seen as the high-frequency branch of zonal flows, is a well-known phenomenon naturally existing in tokamak plasmas\cite{Diamond}. It was experimentally observed in many tokamak devices (see, for examples, Refs. \onlinecite{JIPPT2005,HL2A2006,Lan2008}) and intensively investigated in terms of theoretical analyses\cite{Gaozhe2006,wang,Fu2008,PRL08,Qiu2012,Gaozhe2013,PPCF2013,Qiu2014}and numerical simulations\cite{Xu2008,Ye2013,HW2013,prl2013}. GAM is basically a sort of electostatic perturbation with toroidally symmetrical and poloidally nearly symmetrical structure. That is, the thermal pressure and density perturbations have wave numbers $m = 1$ and $n = 0$, while the polodial Lagrangian perturbation has a poloidal symmetrical structure with $m = 0$. The typical frequency of GAM in the ideal magnetohydrodynamic (MHD) model reads $\omega_s^2 = c_s^2 (2 + q^{- 2})/R^2$, where $q$ is the safety factor and $c_s = (\Gamma (T_e + T_i)/m_i)^{1/2}$ is the sound speed with the ion mass $m_i$, electron and ion temperature $T_e$ and $T_i$, and adiabatic index $\Gamma$. The ideal MHD model yielded $\Gamma = 5/3$ while kinetic models predicted $\omega^2 = (7/4 + \tau ) v_{Ti}^2 /R^2$\cite{Leb1996,JPP2006}, where $v_{Ti} = \sqrt{2 T_i/m_i}$ is the ion thermal velocity and $\tau$ denotes $T_e/T_i$. It was shown that the account of the pressure anisotropy via the parallel ion viscosity exactly recovered the adiabatic indexes obtained in kinetic models\cite{PPCF2012,pla2013}. By using drift kinetic equation, the GAM in the plasma with bi-Maxwellian distribution for ions was investigated and the previous kinetic result was recovered when zeroing the anisotropy\cite{Zhou2010}. Recently, Ren studied the GAM in an anisotropic plasma by using the Chew-Goldberger-Low (CGL) double entropy equations and found that the kinetic result for zero electron temperature was recovered when zeroing the anisotropy\cite{Ren2014}.

On the other hand, owing to the significant applications to the $\vec{E} \times \vec{B}$ shear flow control of anomalous transport and turbulence, the magnitude, radial profile, and evolution of toroidal flow in tokamak plasmas has been an important issue for tokamaks\cite{Iter,JDCallen,PRL2012}. The equilibrium toroidal rotation flow (ETRF) can be on the order of ion thermal velocity \cite{Hazeltine1974,Wong1982,Hinton1985} and has been shown to be important for GAM and attracted much attention since the seminal monograph by Wang\cite{wang}. Wang first investigated the GAM in a toroially rotating tokamak by using the magnetic surface averaged poloidal motion equation to eliminate the fast magnetosonic wave. Later, Wahlberg found the dispersion relation of GAM by solving the Frieman-Rotenberg eigenvalue equation for the Lagrangian perturbation in the presence of ETRF\cite{PRL08} and presented more detailed analysis about the GAM and low-frequency MHD modes in a following paper \onlinecite{Wahlberg2009}. Lakhin and Ilgisonis used the ideal MHD equations in axisymmetric toroidal systems to investigate the GAM by taking into account both ETRF and equilibrium poloidal rotation flows (EPRF)\cite{POP2011}. Recently, Ren studied the GAM in a toroidally rotating tokamak plasma with an arbitrary $\beta$ by following Wang's way\cite{wang} to deal with the poloidal motion equation\cite{HR2012} and then re-studied the effect of $\beta$ on GAMs in the presence of ETRF by considering the harmonics coupling in lieu of magnetic surface averaging manipulation\cite{HR2013}.

The ETRF can increase the frequency of GAM and induce a new low-frequency branch of zonal flows\cite{wang,PRL08}, and the perturbation is always stable in the MHD model. In view of that the GAM has a Landau damping rate $\gamma \propto q^5 e^{- 7 q^2/4}$, it is of great interest to study the collisionless damping of GAM in the presence of ETRF, which is the scope of the present work. We also note that the GAM driven by the energetic particles can be unstable\cite{Fu2008,Qiu2010,Zarzoso2012,prl2013}. Considering that the plasma rotation flow widely exists in tokamak plasmas, even if no external momentum source is injected\cite{Zhou2014}. In that case, the energetic particles may not start to play a role. Then it is of great importance to consider the kinetic effect of ETRF on the GAM. Here with the aid of modified gyro-kinetic (MGK) equation\cite{GK}, we theoretically analyze the effects of arbitrary toroidal rotation on GAM. It is found the Mach number $M$ increases the GAM frequency and remarkably decreases the damping rate. The validity of classical GK (CGK) equation is also discussed and shows that only for zero $\tau$, CGK is identical with MGK. In the case of non-zero $\tau$, CGK yields the same real frequency of GAM as MGK but introduce an instability with a growth rate proportional to $\tau^2 M^3/q$. The rest content is organized as follows. In Section \ref{sec2}, the equilibrium distribution and MGK equation is presented. The dispersion relation of local GAM in a toroidally symmetric tokamak plasma is derived in Section \ref{sec3}. Section \ref{sec4} is devoted to the detailed discussion about the GAM frequency and damping rate. The validity of CGK equation is checked in Section \ref{sec5} and finally, the conclusion is performed in Section \ref{sec6}.

\section{Equilibrium and MGK Equation}
\label{sec2}

We consider a large-aspect-ratio tokamak plasma with a toroidally symmetric magnetic field $\vec{B} = I (\psi) \nabla \zeta + \nabla \zeta \times \nabla \psi$, and work in the $(r, \theta, \zeta)$ coordinate system, where $\psi(r)$ is the magnetic flux, and $\zeta$ and $\theta$ are the toroidal and poloidal angles, respectively. Circular cross-section is assumed in our calculation with $R = R_0 + r \cos \theta$. The subscript $0$ denotes the equilibrium profile and the prefix $\delta$ denotes the perturbed one, but the subscript is omitted and the equilibrium magnetic field is referred to by $\vec{B}$ directly. The equilibrium distribution function for ions with an ETRF is given by \cite{Hinton1985,PJC1987}
\begin{align}
\label{eq1}
F_0^i = n_0(\psi, \theta) (\pi v_{Ti}^2)^{- 3/2} e^{ - (\vec{v} - \vec{u}_0)^2/v^2_{Ti}},
\end{align}
where $\vec{u}_0 = \omega_T(\psi) R^2 \nabla \zeta$ is the toroidal rotational velocity and $n_0$ is the number density of ions,
\begin{align}
n_0(\psi,\theta) = N(\psi) e^{ M^2/(1 + \tau) }.
\end{align}
Here, $M = \omega_T R/v_{Ti}$ is the Mach number. Noting that $T_e$ and $T_i$ are both functions of $\psi$, $M$ depends on the poloidal angle via $R$. As for electrons, the equilibrium distribution function has a standard Maxwellian from, $F_0^e = n_0 (\pi v_{Te}^2)^{- 3/2} e^{- m_e v^2/(2 T_e)}$. Such a distribution requires $\tau \gg m_e/m_i$. That is, the time scale for electron relaxation can be much shorter than the ion time scale\cite{Hinton1985}. The two distribution functions above are the cornerstones of the analysis.

The perturbed distribution function is determined by MGK equation, which reads\cite{GK,CGK,CGK1968}
\begin{align}
\delta F_j = (\partial F_0^j/\partial E ) q_j \delta \phi + [1 - J_0^2(k_r \rho_j) ] \frac{\partial F_0^j}{B \partial \mu} + J_0 \delta h_j,
\end{align}
in which $\delta h_j$ is governed by
\begin{align}
& \bigg[\frac{\partial }{\partial t} +  (w_\parallel \vec{b} + \vec{u}_0 + \vec{v}_D ) \cdot \nabla \bigg] \delta h_j\nonumber\\
= & - q_j J_0 \frac{\partial F_0^j}{\partial E} \bigg( \frac{\partial }{\partial t} + \vec{u}_0 \cdot \nabla \bigg) \delta \phi - J_0 \frac{q_j}{m_j \omega_c^j} \vec{b} \times \nabla \delta \phi \cdot \nabla F_0^j \nonumber\\
+ & J_0 \frac{q_j}{\omega_c^j} \vec{b} \times \nabla \delta \phi \cdot [(w_\parallel \vec{b} + \vec{u}_0) \cdot \nabla \vec{u}_0 + \nabla \vec{u}_0 \cdot (w_\parallel \vec{b} + \vec{u}_0)] \frac{\partial F_0^j}{\partial E}.
\end{align}
Only perturbed electrostatic potential is taken into account, which is justified for electrostatic GAM in a low-$\beta$ plasma. Here, $\mu = \frac{1}{2 B} m_j w_\perp^2$ is the magnetic moment, $E$ is the energy defined in Eq. \eqref{E} below, $k_r$ is the radial wave number of GAM, $\delta \phi$ is the perturbed electrostatic potential, $q_j(= \pm e)$ is the charge of species $j$ ($j = i, e$ for ions and electrons, respectively), $J_0$ is the zeroth-order Bessel function, $\rho_j = w_\perp/\omega_c^j$ is the Larmor radius, $\omega_c^j = q_j B/m_j$ is the gyro frequency with $m_j$ being the mass, $\vec{w} = \vec{v} - \vec{u}_0$ is the particle velocity in the local reference frame moving with the velocity $\vec{u}_0$ relative to the lab frame, and $\vec{v}_D$ is the leading order drift velocity. In the lab frame, electric field can be expanded as $\vec{E} = \vec{E}_{- 1} + \vec{E}_0 + \cdots$, where $\vec{E}_{- 1} = - \vec{u}_0 \times \vec{B}$ and $\vec{E}_0$ is related to $\Phi_0$. Equilibrium analysis\cite{Hinton1985} shows $e \Phi_0 = \frac{m_i \omega_T^2 R^2}{2 (1 + \tau^{- 1})}$. Hence in the local reference frame, particles only feel the potential $\Phi_0$. As a result, the drift velocity can be expressed as
\begin{align}
&\vec{v}_D = [(w_\parallel^2 + w_\perp^2/2)/\omega_c^j] \vec{b} \times \nabla \ln B\nonumber\\
&+ \frac{\vec{b}}{\omega_c^j} \times \bigg[\frac{q_j}{m_j} \nabla \Phi_0 + \vec{u}_0 \cdot \nabla \vec{u}_0 + w_\parallel (\vec{b} \cdot \nabla \vec{u}_0 + \vec{u}_0 \cdot \nabla \vec{b}) \bigg].
\end{align}
Meanwhile, the energy $E$ is defined as
\begin{align}
\label{E}
E = \frac{1}{2} m_j w^2 - \frac{1}{2} m_j u_0^2 + q_j \Phi_0.
\end{align}
As a result, the equilibrium distribution can be arranged as
\begin{align}
F_0^j = N(\psi) (\pi v_{Tj}^2)^{- 3/2} e^{- E/T_j}.
\end{align}

\section{Perturbed Distribution Function And Dispersion Relation}
\label{sec3}
We focus on the ions perturbed distribution function first. Using the properties of $\vec{u}_0$, one can show that $\nabla \vec{u}_0 = \omega_T R (\nabla R \nabla \zeta - \nabla \zeta \nabla R) + R^2 \nabla \omega_T \nabla \zeta$, $\vec{u}_0 \cdot \nabla \vec{b} = \vec{b} \cdot \nabla \vec{u}_0$, and
\begin{align}
\vec{v}_D \cdot \nabla \psi = & \,\frac{I B}{\omega_c^i} (w_\parallel^2 + w_\perp^2/2) \nabla_\parallel \bigg( \frac{1}{B} \bigg) + \frac{\omega_T^2 I}{2 \omega_c^i (1 + \tau)} \nabla_\parallel R^2\nonumber\\
&  + \frac{w_\parallel \omega_T B}{\omega_c^i} \nabla_\parallel R^2,
\end{align}
as well as
\begin{align}
(w_\parallel \vec{b} + \vec{u}_0) \cdot \nabla \vec{u}_0 + \nabla \vec{u}_0 \cdot (w_\parallel \vec{b} + \vec{u}_0) = \bigg( \frac{I w_\parallel}{B} + \omega_T R^2 \bigg) \nabla \omega_T.
\end{align}
In view of the two equations above, MGK equation is reduced to
\begin{align}
\partial_\theta \delta h_i - i n_d^i \sin \theta \delta h_i - i \frac{\omega}{\omega_t^i} \delta h_i = - i J_0 \frac{e \omega}{T_i \omega_t^i} F_0^i \delta \phi.
\end{align}
Here, $\omega_t^i = w_\parallel/(q R)$ is the transit frequency and $n_d^i = k_r \delta_b^i$, where $\delta_b^i$ is the ions orbit width defined as $\frac{1}{\omega_t^i \omega_c^i R} \bigg( w_\parallel^2 + \frac{1}{2} w_\perp^2 + \frac{M^2 v_{Ti}^2}{1 + \tau} + 2 w_\parallel v_{Ti} M \bigg)$. By assuming $\delta \phi = \sum \delta \phi_n e^{i n \theta}$, the disturbed distribution function can be easily solved as
\begin{align}
\label{dis1}
 \delta F_i  & =  - \frac{e}{T_i} F_0^i [1 - J_0^2 (k_r \rho_i)] \delta \phi - \frac{e}{T_i} J_0^2 F_0^i\nonumber\\
& \times \sum_{n, k} i^{n - k} J_{n + l - k} (n_d^i) J_l(n_d^i) \frac{(l - k) \delta \phi_n e^{i k \theta}}{l - k + \omega/\omega_t^i} .
\end{align}
Generally for ions, the finite-Larmor-radius (FLR) effect is taken into account by assuming $k_r \rho_i \thicksim \Delta \ll 1$ and the finite-orbit-width (FOW) effect is also considered by assuming $n_d^i \thicksim \Delta$. While for electrons, we have $k_r \rho_e \simeq 0$ and $n_d^e \simeq 0$. The electron disturbed distribution function is simplified to\cite{Gaozhe2006}
\begin{align}
\label{dis2}
\delta F_e = \frac{e}{T_e} F_0^e \sum_{k \neq 0} \delta \phi_k e^{i k \theta}.
\end{align}

The dispersion relation of GAM is derived by using quasi-neutrality condition, $\delta n_i = \delta n_e$, namely, $\int d^3 w \delta F_i = \int d^3 w \delta F_e$. Inserting $\delta F_j $ into the quasi-neutrality condition will yield the dispersion relation. Before doing that, we need to simplify $\delta F_i$ by cutting off the coupling chains. Considering $\delta \phi_n \thicksim (k_r \rho^i)^n \delta \phi_0$, we take into account only the coupling between $\delta \phi_0 $ and $\delta \phi_{\pm 1}$ by neglecting all high-order harmonics. Here, $\rho^i = v_{Ti}/\omega_c^i$ is the ion Larmor radius. Consequently, we write $\delta F_i = \delta F_i^0 + \delta F_i^{\pm 1} e^{\pm i \theta}$. Eventually, using Eqs. \eqref{dis1} and \eqref{dis2} and the quasi-neutrality condition, $\delta \phi_k$ is obtained as $[1 + \tau (1 + \zeta \mathcal{Z}(\zeta) ) ] \delta \phi_k = - i \frac{1}{2} q k_r \rho^i \tau k \mathcal{G}(k M) \delta \phi_0$ for $k = \pm 1$. For $k = 0$, we find
\begin{align}
\mathcal{S}(M) \delta \phi_0 - \frac{i}{q k_r \rho^i} \mathcal{G}(M) \delta \phi_1 + \frac{i}{q k_r \rho^i} \mathcal{G} (- M) \delta \phi_{- 1} = 0.
\end{align}
Finally, we obtain the following dispersion relation
\begin{align}
\label{dis}
S - \frac{1}{2} \tau \frac{\mathcal{G}^2(M) + \mathcal{G}^2(- M)}{ 1 + \tau (1 + \zeta \mathcal{Z}(\zeta))} = 0,
\end{align}
in which we have denoted
\begin{align}
\label{s}
\mathcal{S}(M) & = \,\frac{1}{q^2} + \frac{3}{2} + \frac{6 + 4 \tau}{1 + \tau} M^2 + \zeta^2 + \frac{\mathcal{Z}(\zeta)}{\zeta} \bigg[ \zeta^4 \nonumber\\
& + \bigg(1 + \frac{6 + 4 \tau}{1 + \tau} M^2 \bigg) \zeta^2 + \frac{1}{2} + \frac{M^2}{1 + \tau} \bigg(1 + \frac{M^2}{1 + \tau} \bigg) \bigg],\\
\label{G}
\mathcal{G}(M) & =  \,\zeta + 2 M + \bigg(\zeta^2 + 2 \zeta M + \frac{1}{2} + \frac{M^2}{1 + \tau} \bigg) \mathcal{Z}(\zeta).
\end{align}
Here, $\zeta$ is defined as $ q \omega R/v_{Ti}$ and $\mathcal{Z} (\zeta) $ is the plasma dispersion function. The previous result\cite{Gao2008} is recovered by zeroing $M$. For convenience of discussion, the Mach number $M$ is assumed to be positive since the dispersion relation above is an even function about $M$. In other words, effects of ETRF on the GAM is independent of the fact that the rotational flow is parallel or antiparallel to the longitude current as predicted in the MHD model\cite{PRL08}.

\section{GAM Frequency And Damping Rate}
\label{sec4}
Explicit analytical solutions to the dispersion relation \eqref{dis} for arbitrary $\zeta$ are difficult to obtain. Here we are restricted to the GAM with $\zeta \gg 1$ to find the asymptotic solution. Since there is $\zeta^2 = q^2 (7/4 + \tau)$ in the non-rotating plasma, $\zeta \gg 1$ requires large safety factor. Hence, analytical results below are expected to be accurate enough only when $q$ is high enough. Now we can asymptotically expand the plasma dispersion function $ \mathcal{Z} (\zeta) = i \sigma \sqrt{\pi} \exp{( - \zeta^2)} - \zeta^{- 1} (1 + \zeta^{- 2}/2  + 3 \zeta^{- 4}/4  + 15 \zeta^{- 6}/8 + \cdots)$. Neglecting all terms of order higher than $\mathcal{O}(\zeta^{- 6})$ leads to the following reduced dispersion relation:
\begin{align}
\label{dis4}
\frac{1}{q^2} - \frac{\mathcal{G}_1}{\zeta^2} - \frac{\mathcal{G}_0}{\zeta^4} + i \sqrt{\pi} \sigma \zeta^3 e^{- \zeta^2} = 0,
\end{align}
in which
\begin{align}
&\mathcal{G}_1 = \frac{7}{4} + \tau + \frac{\tau}{2 q^2} + 4 M^2  + \frac{M^4}{1 + \tau},\\
&\mathcal{G}_0 = \frac{23}{8} + \frac{9 \tau}{8} + \frac{3 \tau}{4 q^2} + 5 M^2 + \frac{M^4}{2 (1 + \tau)},
\end{align}

We now separate $\zeta$ into two parts, $q \Omega_K + i q \gamma_d$ with $\Omega_K \gg \gamma_d$. Here, $\Omega_K$ is the normalized frequency of GAM and $\gamma_d$ is the damping rate. $\Omega_K$ is then determined by
\begin{align}
\label{fre}
\Omega_K^2 = \frac{\mathcal{G}_1}{2} + \sqrt{\frac{\mathcal{G}_1^2}{4} + \frac{\mathcal{G}_0}{q^2} },
\end{align}
and the imaginary part of \eqref{dis4} gives the damping rate of GAM in the presence of ETRF as
\begin{align}
\label{damp}
\gamma_d = - \frac{\sqrt{\pi} q^5 \Omega_K^6 }{2 \sqrt{\mathcal{G}_1^2 + 4 \mathcal{G}_0/q^2}} e^{- q^2 \Omega_K^2},
\end{align}
where $\sigma = 1$ is adopted due to the factor that $\gamma_d \ll \Omega_K$. The two equations above are the major results representing the effects of ETRF on the GAM.

Before further discussion about Eqs. \eqref{fre} and \eqref{damp}, let us pay attention to the case of $M = 0$. The original dispersion relation is reduced to the one in Ref. \onlinecite{Gao2008} when zeroing $M$, while the simplified dispersion relation \eqref{dis4} is different with previous results. To illustrate this difference clearly, previous results are written here [see, for example, Eq. (7) in Ref. \onlinecite{Gao2008} and Eq. (30) in Ref. \onlinecite{Wang2011}] as
\begin{align}
\frac{1}{q^2} & - \frac{1}{\zeta^2} \bigg( \frac{7}{4} + \tau \bigg) - \frac{1}{\zeta^4} \bigg( \frac{23}{8} + 2 \tau + \frac{\tau^2}{2} \bigg)\nonumber\\
  & + i \sqrt{\pi} \zeta^3 e^{- \zeta^2} = 0.
\end{align}
The coefficients of $\zeta^{- 2}$ and $\zeta^{- 4}$ terms are different with their analogs in Eq. \eqref{dis4}. These tiny differences are induced by different asymptotic expansion. To obtain a more accurate expression of simplified dispersion relation, we multiplied Eq. \eqref{dis} by $1 + \tau (1 + \zeta \mathcal{Z}(\zeta))$ and then kept the terms to the leading order. However, on the order of $\mathcal{O}(1/q^2)$, Eq. \eqref{dis4} yields the GAM frequency in non-rotating plasma as
\begin{align}
\Omega_G^2 = \bigg( \frac{7}{4} + \tau \bigg) \bigg[1 + \frac{46 + 32 \tau + 8 \tau^2}{(7 + 4 \tau)^2 q^2} \bigg],
\end{align}
which is identical with the one in Ref. \onlinecite{Gao2008}.

Now we take into account the presence of Mach number. Eq. \eqref{fre} illustrates the dependence of GAM frequency on the safety factor $q$, temperature ratio $\tau$, and the Mach number $M$. For general Mach number $M \thicksim \mathcal{O}(1)$, by ignoring terms proportional to $1/q^2$, GAM frequency is reduced to
\begin{align}
\Omega_K^2 = \frac{7}{4} + \tau + 4 M^2 + \frac{M^4}{1 + \tau}.
\end{align}
While the MHD model yields $(\Omega_K^{\textrm{MHD}})^2 = \Gamma (1 + \tau) + 4 M^2 + \frac{M^4}{1 + \tau}$ according to Ref. \onlinecite{PRL08}. The coefficients of $M^2$ and $M^4$ derived in the GK model are identical with their MHD analogs. The only difference between the MHD and GK results lies on the adiabatic index $\Gamma(1 + \tau)$ and $\frac{7}{4} + \tau$. Obviously, the GAM frequency is increased by the increasing Mach number as shown in the MHD framework\cite{PRL08}.

%

\begin{figure}[h]
\setlength{\unitlength}{0.5cm}
\begin{center}
\begin{minipage}[t]{8cm}
\includegraphics[width=8cm]{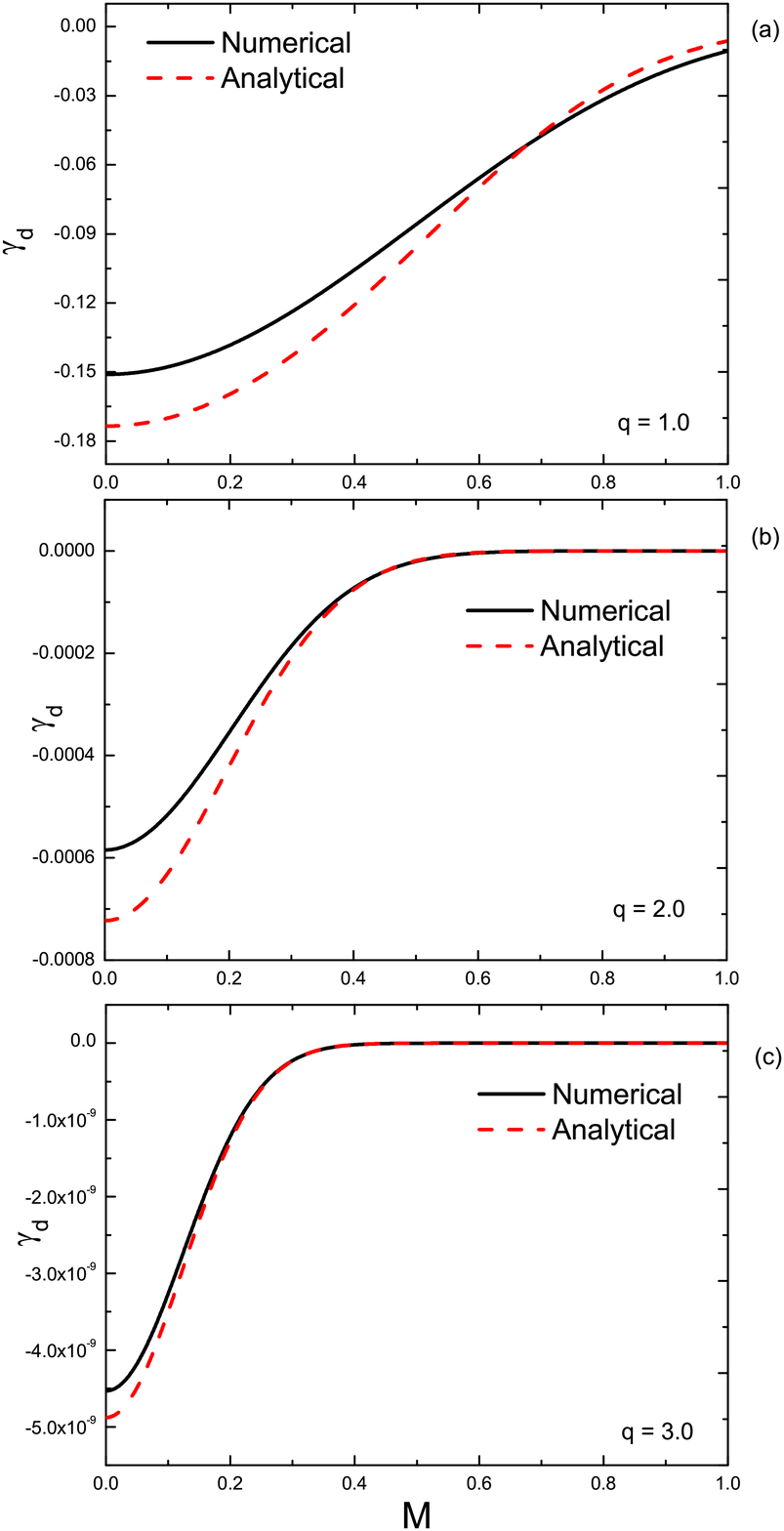}
\caption{The dependence of Landau damping rate on $M$ according to the exact numerical solution to the dispersion relation \eqref{dis} (solid curves) and to the analytical expression \eqref{damp}, respectively, for $\tau = 1$ and fixed safety factor, $q = 1$ (a), $q = 2$ (b), and $q = 3$ (c).  }
\label{fig1}
\end{minipage}
\end{center}
\end{figure}

The toroidal Mach number is shown to decrease the damping rate by increasing the frequency. The dependence of the Landau damping rate on the Mach number is plotted in Fig. \ref{fig1}. One can see that the damping rate is dramatically decreased by the increased $M$. That is, the Mach number tends to increase the GAM frequency and destabilize the GAM by diminishing the damping rate. Besides, according to Fig. \ref{fig1}, the analytical result \eqref{damp} differs from the exact numerical result for $q = 1$ or for $q = 2$ when $M < 0.3$. For $q > 2$ or large $M$, Eq. \eqref{damp} agrees well with the numerical result. It is, of course, not surprising since Eq. \eqref{damp} is valid for large $\zeta$, which requires large $q$ or large $M$.

\section{Dispersion Relation in CGK Model}
\label{sec5}
It is argued that CGK model is suitable for describing the toroidally rotating plasmas or not. For convenience of discussion, we still focus on ions first. CGK equation gives the ions perturbed distribution function as $\delta F = e (\partial F_0^i/\partial U) \delta \phi + (1 - J_0^2) \frac{\partial F_0^i}{B \partial \mu} + J_0 \delta h$ with $\delta h$ determined by\cite{CGK,CGK1968}
\begin{align}
\label{gk2}
(- i \omega + \vec{V}_0 \cdot \nabla) \delta h = i e J_0 Q \delta \phi.
\end{align}
Here, $U = \frac{1}{2} m_i v^2 + e \Phi$ is the total energy and $Q = \omega \partial F_0^i/\partial U + \vec{k} \times \vec{b} \cdot \nabla F_0^i/(e B)$. The equilibrium potential reads $\Phi = \Phi_{- 1} + \frac{m_i \omega_T^2 R^2}{2 e (1 + \tau^{- 1})}$ in the lab reference frame with $\Phi_{- 1}(\psi) = - \int \omega_T d \psi$. As a result, the equilibrium electrostatic field is $\vec{E}_0 = \omega_T \nabla \psi - \nabla \frac{m_i \omega_T^2 R^2}{2 e (1 + \tau)}$. The lowest order velocity is $\vec{V}_0 = v_\parallel \vec{b} + [(v_\parallel^2 + v_\perp^2/2)/\omega_c^j] \vec{b} \times \nabla \ln B + \vec{v}_E$. It is easy to obtain that $\vec{v}_E \cdot \nabla \delta h = - \omega_T R (\vec{b} \cdot \nabla) \delta h + i k_r \frac{\omega_T^2 R \sin \theta}{\omega_c^i (1 + \tau^{- 1})} \delta h$.

Poloidal asymmetry of $F_0^i$ needs to be taken into account due to the definition of $U$. Noting that the spacial dependence of $F_0^i$ should be calculated at fixed energy $U$ and magnetic moment $\mu$, we rearrange $F_0^i$ as
\begin{align}
\label{fenbu}
F_0^i = \overline{N}(\psi) e^{- \frac{U}{T_i} + \frac{m \omega_T R u_\parallel}{T_i}} (\pi v_{Ti}^2)^{-3/2}.
\end{align}
We can find
\begin{align}
\partial_\theta F_0^i = - 2 M \frac{r v_{Ti} }{R v_\parallel} F_0^i \bigg( \frac{v_\parallel^2}{v_{Ti}^2} + \frac{v_\perp^2}{2 v_{Ti}^2} - \frac{\tau M^2}{1 + \tau}\bigg) \sin \theta.
\end{align}
That is to say, due to the outward shift of ions, the ions distribution function becomes poloidally asymmetric under fixed energy $U$ and $\mu$. As a result, we can reexpress $Q$ as $Q^c - \mathcal{N}_i \sin \theta$ with $Q^c = \omega \partial F_0^i/\partial U$ and
\begin{align}
\mathcal{N}_i = - \frac{k_r v_d}{T_i} \frac{ M v_{Ti} }{v_\parallel } F_0^i.
\end{align}
The CGK equation \eqref{gk2} is rewritten as
\begin{align}
\partial_\theta \delta h - i n_d \sin \theta \delta h  - i \frac{\omega}{\omega_t} \delta h = i e J_0 \frac{Q^c}{\omega_t} \delta \Phi - i e J_0 \frac{\mathcal{N}_i}{\omega_t} \sin \theta \delta \phi.
\end{align}
Here, $n_d = k_r v_d/\omega_t$, $v_d = \bigg(v_\parallel^2 + \frac{1}{2} v_\perp^2 - \frac{M^2 v_{Ti}^2}{1 + \tau^{- 1}} \bigg)/(R \omega_c^i) $ is the zeroth-order radial drift velocity, and $\omega_t = (v_\parallel/R - \omega_T)/q$ is the modified transit frequency. It should be noted that $\omega_t = \omega_t^i$ and $v_d = \delta_b^i \omega_t^i$ by recalling $v_\perp = w_\perp$ and $v_\parallel = w_\parallel + \omega_T R$. After some algebraic manipulation, the perturbed distribution function of passing ions is analytically given as
\begin{align}
\label{disn}
\delta F_i = & \,e \bigg(\frac{\partial F_0^i}{\partial U} + \frac{\partial F_0^i}{B \partial \mu} \bigg) [ 1 - J_0^2 (k_r \rho_i) ] \delta \phi + J_0^2 e \frac{\partial F_0^i}{\partial U} \nonumber\\
 \times & \sum i^{n - k} J_{n + l - k} (n_d) J_l (n_d) \frac{l - k}{l - k + \omega/\omega_t} \delta \phi_n e^{i k \theta}\nonumber\\
 &+ J_0^2 \frac{e \mathcal{N}_i}{2 i \omega} \bigg[e^{i \theta} \delta \phi - e^{- i \theta} \delta \phi \nonumber\\
+ 2 & \sum i^{n - k - 1} \frac{(l - k)(n + l - k) }{(l - k + \omega/\omega_t) n_d } J_{n + l - k} J_l \delta \phi_n e^{i k \theta}.
\end{align}
For zero $M$, the total energy $U$ is replaced by the kinetic one $E = \frac{1}{2} m_i v^2$. One then can find $\mathcal{N}_i = 0$ and the perturbed distribution function above is reduced to the previous one in a non-rotating plasma. As for electrons, there is $\mathcal{N}_e = 0$. Neglecting the FLR and FOW effects, the electron disturbed distribution function is the same as Eq. \eqref{dis2}.

According to Eq. \eqref{fenbu}, one has
\begin{align}
&\frac{\partial F_0^i}{\partial U} = - \frac{F_0^i}{T_i} \bigg( 1 - \frac{M v_{Ti}}{v_\parallel} \bigg),\\
&\frac{\partial F_0^i}{B \partial \mu} = - \frac{M v_{Ti}}{T_i v_\parallel} F_0^i.
\end{align}
In view of $\frac{\partial F_0^e}{\partial U} = - \frac{F_0^e}{T_e}$ and after some similar manipulation, we obtain the following dispersion relation
\begin{align}
\label{dis5}
S(M) - \frac{1}{2} \tau \bigg [ \frac{\mathcal{G}(M) \mathcal{D}(M)}{\mathcal{A}(M)} + \frac{\mathcal{G}(- M) \mathcal{D}(- M)}{\mathcal{A}(- M)} \bigg] = 0,
\end{align}
in which $\mathcal{S}(M)$ and $\mathcal{G}(M)$ are defined in Eqs. \eqref{s} and \eqref{G}, respectively, and $\mathcal{A}(M)$ and $\mathcal{D}(M)$ are defined as
\begin{align}
\mathcal{A}(M) = &\,1 + \tau \bigg(1 + \frac{\zeta^2 \mathcal{Z}(\zeta) + M^2 \mathcal{Z}(M)}{\zeta + M} \bigg),\\
\mathcal{D}(M) = &\,(\zeta + M) (1 + \zeta \mathcal{Z}(\zeta) ) - \frac{M^2}{\zeta}\nonumber\\
 & + \frac{\zeta^2 \mathcal{Z}(\zeta) + M^2 \mathcal{Z}(M)}{\zeta (\zeta + M)} \bigg( \frac{1}{2} - \frac{\tau M^2}{1 + \tau} \bigg).
\end{align}
Although just like Eq. \eqref{dis}, this dispersion relation still can reproduce the classical one in the non-rotating case, Eqs. \eqref{dis5} and \eqref{dis} are not identical with each other. Only for zero electron temperature, CGK is exactly identical with MGK. For simplicity of discussion, let us restricted ourselves to the small Mach number case, i.e., the case of $M \ll 1$ since $\mathcal{Z}(M)$ is presented in Eq. \eqref{dis5}. Then $\mathcal{Z}(M)$ can be asymptotically expanded as $i \sqrt{\pi} \exp{(- M^2)} - 2 M (1 - 2 M^2/3 + \cdots)$. Thereby, we can suppose the following ordering relation: $M \thicksim \epsilon$ and $1/q \thicksim \epsilon$. Neglecting all terms of order higher than $\epsilon^4$ in Eq. \eqref{dis5} leads to the following reduced dispersion relation:
\begin{align}
\label{disn4}
\frac{1}{q^2} & - \frac{\mathcal{G}_1'}{\zeta^2} - \frac{\mathcal{G}'_0}{\zeta^4} + i \sqrt{\pi} \bigg[\sigma \zeta^3 e^{- \zeta^2} \nonumber\\
& + M^3 e^{- M^2} \frac{\tau}{\zeta^4} \bigg( \frac{7}{2} + 5 \tau - \frac{2 \zeta^2}{q^2} (1 + \tau) \bigg) \bigg] = 0,
\end{align}
in which
\begin{align}
&\mathcal{G}'_1 = \frac{7}{4} + \tau + \frac{\tau}{q^2} + 4 M^2 \nonumber\\
& \qquad \qquad  + \frac{M^4}{1 + \tau} + \boxed{\frac{M^2}{q^2} (1 + \tau)^2},\\
&\mathcal{G}'_0 = \frac{23}{8} +¡¡\frac{\tau}{4} (1 - 2 \tau) + \frac{6 - \tau}{4 q^2} \tau \nonumber\\
 & \qquad \qquad + \boxed{\frac{M^2}{4} (13 - 30 \tau - 15 \tau^2)},
\end{align}
It should be specifically pointed out that due to coupling effect of $\delta \phi_{\pm 1}$ induced by the nonzero $\tau$ and $M$, we need multiply the dispersion relation \eqref{dis5} by $\zeta^2 - M^2$ to asymptotically expand the equation. As a result, the boxed term in the coefficients above is not accurate, or more precisely, can not reduce to the ones in the case of zero $\tau$ by zeroing $\tau$ directly. While Eq. \eqref{dis} can reproduce the dispersion relation with zero $\tau$ by letting $\tau = 0$ directly.

Albeit there is tiny difference between Eq. \eqref{fre} and the real part of Eq. \eqref{disn4}, the formula above yields the real frequency of GAM to the first order as
\begin{align}
\Omega_K^2 = \bigg( \frac{7}{4} + \tau \bigg) \bigg[1 + \frac{46 + 32 \tau + 8 \tau^2}{(7 + 4 \tau)^2 q^2} \bigg] + 4 M^2 + \frac{M^4}{1 + \tau}
\end{align}
Apparently, this frequency is the same as the MGK result. That is to say, from the aspect of GAM frequency, CGK is the same valid and accurate as MGK for small Mach number. When $M$ is on the order of unit, the CGK dispersion relation \eqref{disn4} needs numerical evaluation, which will not be presented here. While we pay attention to the damping rate, Eq. \eqref{disn4} shows remarkable difference with Eq. \eqref{damp}. Due to the exponential decay of Landau damping rate, the second term can easily exceed the first term in the square brackets of Eq. \eqref{disn4}. For example, when $M > 0.0046$ for $\tau = 1$ and $q = 3$, the second term becomes greater than the first term. Substituting $\zeta^2 \simeq q^2 (7/4 + \tau)$, an instability comes into being with a growth rate
\begin{align}
\gamma_g = 2 \sqrt{\pi} (1 + 4 \tau) (7 + 4 \tau)^{- \frac{3}{2}} \frac{M^3}{q} \tau^2.
\end{align}
Due to the fact that $\gamma_d$ in Eq. \eqref{damp} and $\gamma_g$ above are both much less than $\Omega_K$ while CGK gives the same $\Omega_K$ as MGK, it can not be analytically determined here which of CGK and MGK is more reliable in the case of small Mach number. Experimental observations related to the GAM in the presence of torodial rotation may tell the answer.

\section{Conclusion}
\label{sec6}
Geodesic acoustic mode (GAM) in a toroidally rotating tokamak is investigated by using modified gyro-kinetic (MGK) equation. The equilibrium rotational flow is assumed to be along the toroidal direction since the zeroth-order poloidal flow is forbidden due to the neoclassical constraints\cite{Hinton1985}. The perturbed distribution function is analytically obtained from the GK equation with $m = n = 0$ for the perturbed potential $\delta \phi$. By focusing on the passing ions and assuming small drift orbit radius, the general dispersion relation of GAM is derived, in which the sonic Mach number $M$ represents effects of toroidal rotation and $\zeta$ is the frequency normalized by $v_{Ti}/R$. For large safety factor, the GAM frequency is $\Omega_K^2 = \frac{7}{4} + \tau + 4 M^2 + \frac{M^4}{1 + \tau}$ to the leading order. It is shown that the coefficients of $M^2$ and $M^4$ in the kinetic result are the same as the ones in the MHD result. Torodial Mach number $M$ increases the GAM frequency as predicted by the MHD model\cite{PRL08}, and dramatically decreases the Landau damping rate. The applicability of classical GK (CGK) equation for small Mach number, in which $F_0 = F_0(\vec{X}, U, \mu)$ and $U$ is defined as $\frac{1}{2} m v^2 + e \Phi$, is also checked. It is found that CGK is identical with MGK for $\tau = 0$. In the case of non-zero $\tau$, only the real frequencies of GAM obtained in both models are the same. The collisionless damping term is exceeded and an instability comes into being with a grow rate proportional to $\tau^2 M^3/q$.

\begin{acknowledgments}
This work was supported by the China National Magnetic Confinement Fusion Science Program under Grant Nos. 2015GB120005 and 2013GB112011, and the National Natural Science Foundation of China No. 11275260.

\end{acknowledgments}

\end{document}